\documentclass[conference]{IEEEtran}
\usepackage{amssymb}
\usepackage{color}
\usepackage{multirow}
\usepackage{mathrsfs}
\newtheorem{theorem}{Theorem}

\newtheorem{proposition}{Proposition}
\newtheorem{corollary}{Corollary}
\newtheorem{remark}{Remark}
\newtheorem{definition}{Definition}
\newtheorem{conjecture}{Conjecture}

\usepackage{cite}
\ifCLASSINFOpdf
  \usepackage[pdftex]{graphicx}
  \graphicspath{{../pdf/}{../jpeg/}}
  \DeclareGraphicsExtensions{.pdf,.jpeg,.png}
\else
  \usepackage[dvips]{graphicx}
  \graphicspath{{../eps/}}
  \DeclareGraphicsExtensions{.eps}
\fi
\usepackage[cmex10]{amsmath}
\usepackage{array}
\usepackage[tight,footnotesize]{subfigure}
\usepackage{stfloats}

\usepackage{arydshln}
\usepackage{rotating}
\usepackage{microtype}
\def\gap{.5ex}
\abovedisplayskip\gap
\belowdisplayskip\gap
\abovedisplayshortskip\gap
\belowdisplayshortskip\gap

\IEEEoverridecommandlockouts

\usepackage{epstopdf}
\begin{document}
\title{A Unified Inner Bound for the Two-Receiver Memoryless Broadcast Channel with Channel State and Message Side Information}

\author{\IEEEauthorblockN{Behzad Asadi, Lawrence Ong, and Sarah J.\ Johnson}
	\IEEEauthorblockA{School of Electrical Engineering and Computer Science, The University of Newcastle, Newcastle, Australia}
	Email:{ behzad.asadi@uon.edu.au, lawrence.ong@cantab.net, sarah.johnson@newcastle.edu.au}
}

\maketitle

\begin{abstract}
We consider the two-receiver memoryless broadcast channel with states where each receiver requests both common and private messages, and may know part of the private message requested by the other receiver as receiver message side information (RMSI). We address two categories of the channel (i) channel with states known causally to the transmitter, and (ii) channel with states known non-causally to the transmitter. Starting with the channel without RMSI, we first propose a transmission scheme and derive an inner bound for the causal category. We then unify our inner bound for the causal category and the best-known inner bound for the non-causal category, although their transmission schemes are different. Moving on to the channel with RMSI, we first apply a pre-coding to the transmission schemes of the causal and non-causal categories without RMSI. We then derive a unified inner bound as a result of having a unified inner bound when there is no RMSI, and applying the same pre-coding to both categories. We show that our inner bound is tight for some new cases as well as the cases whose capacity region was known previously.
\end{abstract}
\IEEEpeerreviewmaketitle

\section{Introduction}
Communication over wireless channels motivates the study of the broadcast channel~\cite{BC}, a common situation where a transmitter sends a number of messages to multiple receivers via a shared medium. 

In some scenarios, the messages to be sent by the transmitter may already be present in
parts at the receivers, referred to as \textit{receiver message side information} (RMSI). This form of side information appears in, for example, the downlink phase of applications modeled by the multi-way relay channel~\cite{MWRCFullExchange}. A broadcast channel may be time-varying due to, for example, fading or interference. The \textit{channel state} capturing this variation over time may be available causally or non-causally at each node (each of the transmitter and receivers) as side information~\cite{DegradedBCCausalNonCausal}. 

It is well-known that proper use of side information available at each node may increase the
transmission rates over the channel. With this as motivation, we investigate the capacity region of the two-receiver memoryless broadcast channel with states and RMSI under two categories: (i) channel with causal channel state side information at the transmitter (CSIT) including the cases where the channel state may also be available at each receiver\footnote{It does not make a difference whether the channel state is available causally or non-causally at a receiver; this is due to block decoding, where the receiver decodes its requested message(s) at the end of the channel-output sequence.}, and (ii) channel with non-causal CSIT including the cases where the channel state may also be available at each receiver. 

\subsection{Existing Results}
Most of the results considering RMSI are for the two-receiver memoryless broadcast channel without state~\cite{TwoReceiverwithoutState, TwoReceiversemideterministic}; our inner bound~\cite{TwoReceiverwithoutState}, achieved using Marton coding~\cite[p. 208]{NITBook}, superposition coding, and our proposed pre-coding, is tight for all the cases whose capacity region is known. For the broadcast channel with states, existing capacity results are as follows.

\subsubsection{With Causal CSIT and RMSI} To the best of our knowledge, there does not exist any work in this category.

\subsubsection{With Non-causal CSIT and RMSI} Under this category, Oechtering and Skoglund~\cite{BCNonCasualandSI0} established the capacity region of the memoryless channel where the channel state is available non-causally at the transmitter and one of the receivers; they considered complementary RMSI where each receiver knows all the messages requested by the other receiver as side information. Xin et~al.~\cite{BCNonCasualandSI1} derived inner and outer bounds for the Gaussian scalar, and Gaussian vector broadcast channel where the channel state is available non-causally at only the transmitter, and the RMSI is complementary (the capacity region of the considered cases remained unknown). Song et al.~\cite{BCNonCasualandSI2} investigated the capacity region of the degraded broadcast channel where the channel state is available non-causally at only the transmitter; they considered (i) complementary RMSI, and (ii) where the weaker receiver knows fully the requested message of the stronger receiver and the stronger receiver has no message side information (the capacity region of the considered cases remained unknown).

\subsection{Contributions}\label{Sec:mainresults}

\begin{figure}[t]
	\centering
	\includegraphics[width=0.45\textwidth]{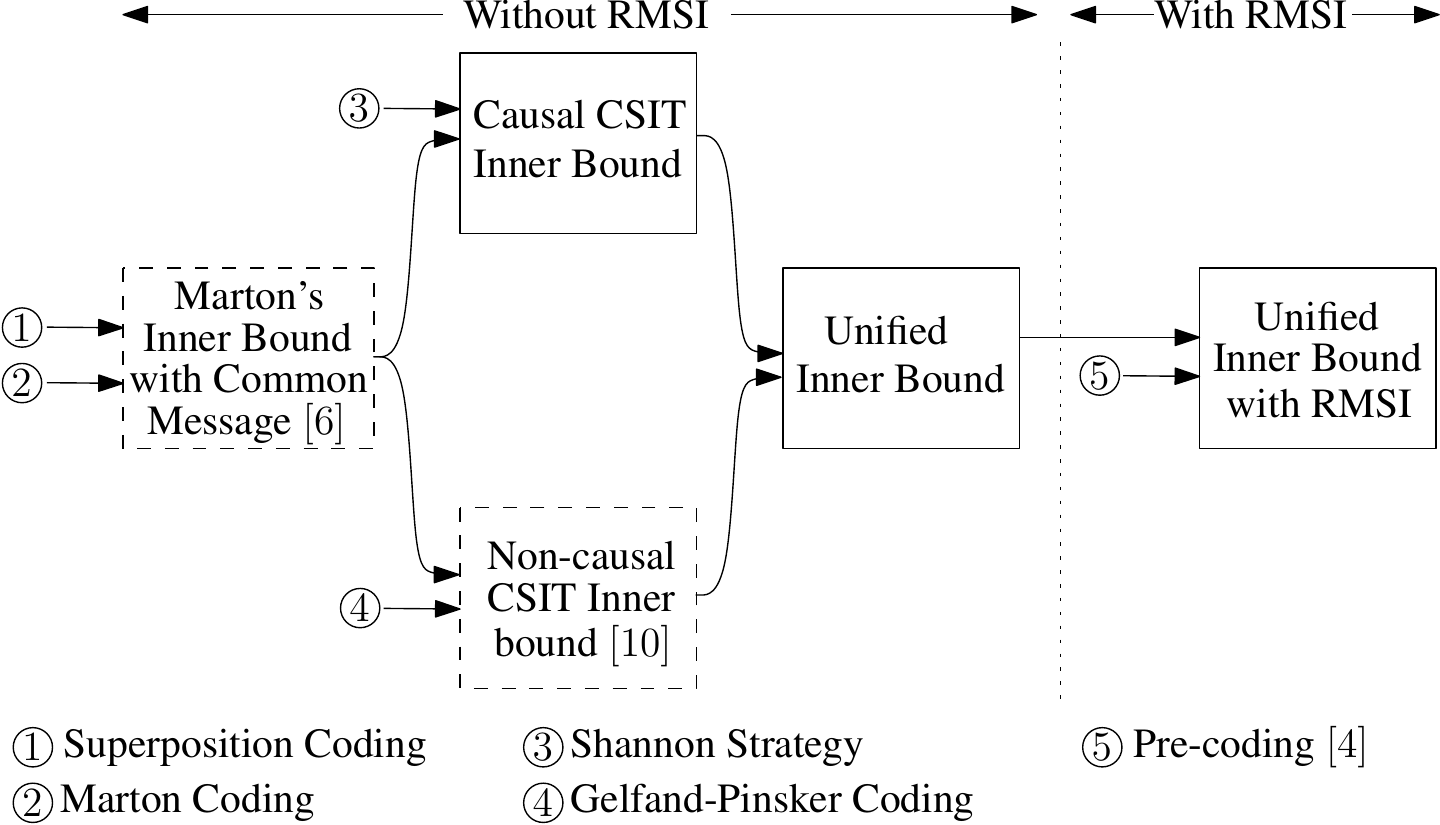}
	\vskip-5pt
	\caption{The steps to derive our unified inner bound for the causal and non-causal categories with RMSI (the rightmost rectangle); each rectangle represents one step, and is labeled by its output. The arrows provide the key techniques to complete each step. The rectangles with solid sides are the steps completed in this work.} 
	\vskip-12pt
	\label{Fig:StepDiagram}
\end{figure}

We investigate the capacity region of the two-receiver memoryless broadcast channel with states and RMSI. We consider the general message setup that includes all possible message requests and RMSI configurations, i.e., each receiver (i) has both common and private-message requests, and (ii) knows part of the private message requested by the other receiver as side information. We derive a unified inner bound that covers both the causal and non-causal categories with RMSI. The steps to derive our unified inner bound are shown in Fig.~\ref{Fig:StepDiagram}, in which rectangles with solid sides represent new bounds established in this work. Here, we briefly explain the steps.

\underline{\textit{Step} 1:} We first propose a transmission scheme and derive a general inner bound for the causal category without RMSI. We use Marton coding, superposition coding, and Shannon strategy~\cite[p. 176]{NITBook} to construct the scheme. This inner bound is tight for all the cases with causal CSIT whose capacity region is known~\cite{DegradedBCCausalNonCausal},~\cite[p.\ 184]{NITBook}. 

\underline{\textit{Step} 2:} We then unify our inner bound for the causal category without RMSI, and the best-known inner bound for the non-causal category without RMSI~\cite[Theorem 2]{MUNonCausal} achieved using Marton coding, superposition coding, and Gelfand-Pinsker coding~\cite[p.\ 180]{NITBook}. This result is analogous to the work of Jafar~\cite{PtPUnifiedCausalNonCausal} in which a unified capacity-region expression is provided for the point-to-point channel with causal CSIT and the point-to-point channel with non-causal CSIT. Clearly, the capacity region of a non-causal case is larger than or equal to the capacity region of the corresponding causal case (where only the transmitter knows the channel state causally instead of non-causally, and the knowledge of the receivers about the channel state is the same in both cases). This relationship is not necessarily true for their inner bounds especially when one transmission scheme is not a special case of the other. One of the advantages of having a unified inner bound is that it allows us to show that the best-known inner bound~\cite[Theorem 2]{MUNonCausal} for a non-causal case is larger than or equal to the best inner bound (our inner bound) for the corresponding causal case.

\underline{\textit{Step} 3:} Moving on to the channel with RMSI,  we use a pre-coding in order to take the RMSI into account; this pre-coding was proposd in our previous work for the channel without state, with RMSI~\cite{TwoReceiverwithoutState}. We use this pre-coding in conjunction with the schemes achieving the best inner bounds for the two categories without RMSI. We finally derive a unified inner bound that covers both the causal and non-causal categories with RMSI. This inner bound reduces to the unified inner bound without RMSI by setting some parameters to zero.

\underline{\textit{Capacity Results} :} Using our inner bound, we establish the following new capacity results for the memoryless broadcast channel with RMSI.

\subsubsection{With Causal CSIT and RMSI} For causal cases, we show that our inner bound establishes the capacity region of the degraded broadcast channel where the channel state is available causally at (i) only the transmitter, (ii) the transmitter and the non-degraded receiver, or (iii) the transmitter and both receivers.

\subsubsection{With Non-causal CSIT and RMSI} For non-causal cases, we show that our inner bound establishes the capacity region of the degraded broadcast channel where the channel state is available non-causally at (i) the transmitter and the non-degraded receiver, or (ii) the transmitter and both receivers.

\section{System Model}\label{Section:SystemModel}
\begin{figure}[t]
	\centering
	\includegraphics[width=0.45\textwidth]{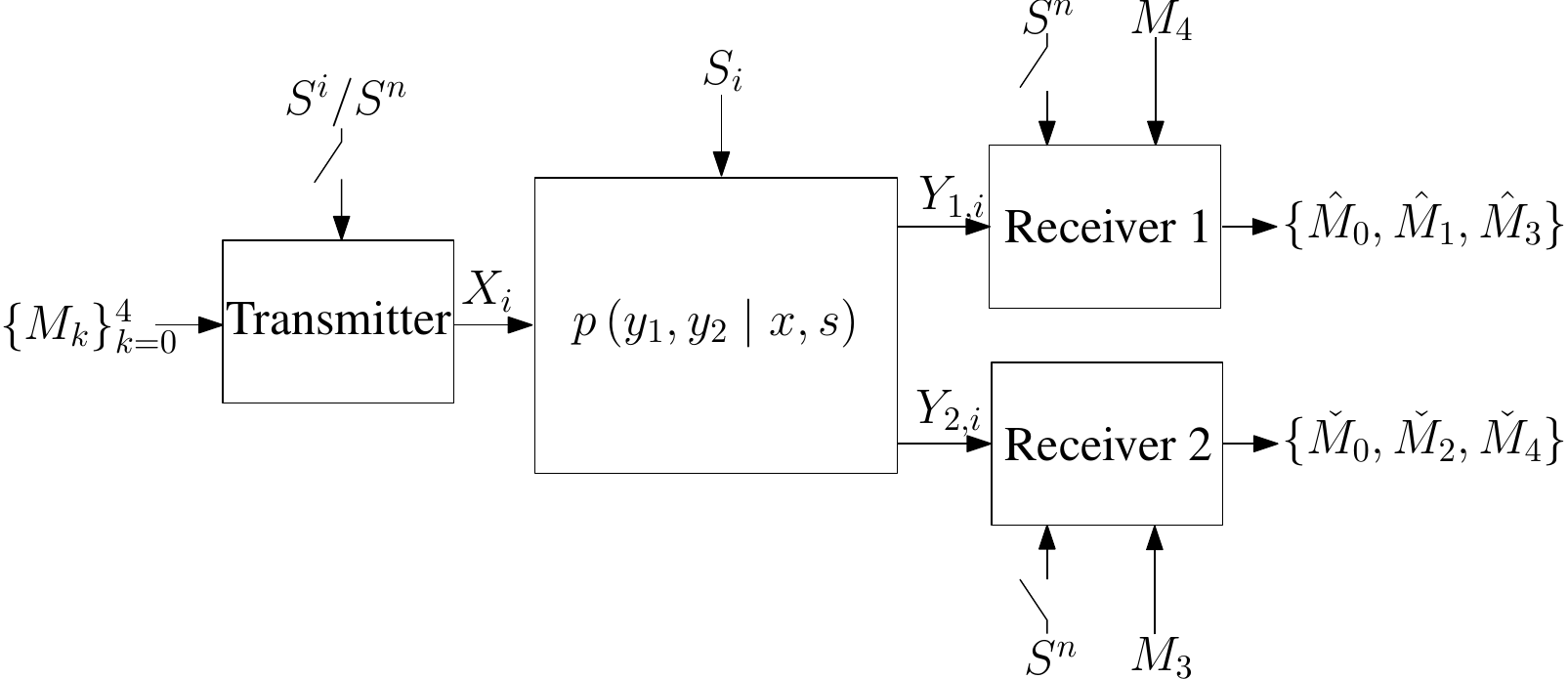}
	\vskip-10pt
	\caption{The two-receiver memoryless broadcast channel $p(y_1,y_2\mid x,s)p(s)$ with i.i.d. states and RMSI. The channel state may be available either causally or non-causally at each of the transmitter and receivers. $\{M_i\}_{i=0}^4$ is the set of independent messages sent by the transmitter. $M_4$ and $M_3$ are the messages known a priori to receivers~1 and~2 respectively. $\{M_0,M_1,M_3\}$ and $\{M_0,M_2,M_4\}$ are the set of messages requested by receivers~1 and~2 respectively. $\hat{M}_i,\;\,i=0,1,3$, is the decoded $M_i$ at receiver 1, and $\check{M}_i,\;\,i=0,2,4$, is the decoded $M_i$ at receiver 2.} 
	\vskip-12pt
	\label{Fig:DM-BCwithState}
\end{figure}

We consider the two-receiver memoryless broadcast channel with independent and identically distributed (i.i.d.) states $p_{_{Y_1,Y_2\mid X,S}}(y_1,y_2\hskip-2pt\mid\hskip-2pt x,s)p_{_{S}}(s)$, depicted in Fig. \ref{Fig:DM-BCwithState}, where $X\in\mathcal{X}$ is the channel input, $Y_1\in\mathcal{Y}_1$ and $Y_2\in\mathcal{Y}_2$ are the channel outputs, and $S\in\mathcal{S}$ is the channel state. Considering $n$ uses of the channel, $X^{n}=\left(X_1,X_2,\ldots,X_n\right)$ is the transmitted codeword, and $Y_{i}^{n}=\left(Y_{i,1},Y_{i,2},\ldots,Y_{i,n}\right),\;\,i=1,2$, is the channel-output sequence at receiver $i$. 


The source messages $\{M_i\}_{i=0}^4$ are independent, and $M_i$ is uniformly distributed over the set $\mathcal{M}_i=\{1,2,\ldots,2^{nR_i}\}$, i.e., transmitted at rate $R_i$ bits per channel use. $\{M_0,M_1,M_3\}$ and $\{M_0,M_2,M_4\}$ are the set of messages requested by receivers~1 and~2 respectively. $M_4$ and $M_3$ are the messages known a priori to receivers~1 and~2 respectively. For receiver~1, $M_1$ is the part of the private-message request which is not known a~priori to the other receiver, and $M_3$ is the part which is known. For receiver~2, these are $M_2$ and $M_4$ respectively.

The channel without state is a special case of our channel model by considering $\mathcal{S}=\{0\}$, i.e., $p_{_S}\hspace{-2pt}\left(0\right)=1$; this implies that all the transmitter and receivers know that the channel state is equal to zero at all channel uses. The channel without RMSI is also a special case of our channel model by considering $(M_3,M_4)=(0,0)$.

A $\left(2^{nR_0},2^{nR_1},2^{nR_2},2^{nR_3},2^{nR_4},n\right)$ causal code for the channel consists of a sequence of maps for the encoding,
\begin{align*}
f_j\hskip-1pt:\hskip-1pt\mathcal{M}_0\hskip-1pt\times\hskip-1pt \mathcal{M}_1\hskip-1pt \times\hskip-1pt \mathcal{M}_2\hskip-1pt \times\hskip-1pt \mathcal{M}_3\hskip-1pt \times\hskip-1pt \mathcal{M}_4\hskip-1pt \times\hskip-1pt {\mathcal{S}}^j\hskip-1pt\rightarrow\hskip-1pt \mathcal{X},\;j=1,2,\cdots,n,
\end{align*}
where $\times$ denotes the Cartesian product, and ${\mathcal{S}}^{j}$ denotes the $j$-fold Cartesian product of ${\mathcal{S}}$, i.e., $X_j=f_j(M_0,M_1,M_2,M_3,M_4,S^j)$ where $S^j=(S_1,S_2,\ldots,S_j)$. This code also consists of two decoding functions,
\begin{align*}
g_1:\tilde{\mathcal{Y}}_1^{n}\times \mathcal{M}_4 \rightarrow \mathcal{M}_0\times \mathcal{M}_1 \times \mathcal{M}_3,\\
g_2:\tilde{\mathcal{Y}}_2^{n}\times \mathcal{M}_3 \rightarrow \mathcal{M}_0\times \mathcal{M}_2 \times \mathcal{M}_4,
\end{align*}
where $\tilde{\mathcal{Y}}_i=\mathcal{Y}_i\times\mathcal{S},\;\;i=1,2$, if the channel state is available at receiver~$i$, and $\tilde{\mathcal{Y}}_i=\mathcal{Y}_i$ otherwise. $(\hat{M}_0,\hat{M}_1,\hat{M}_3)=g_1(\tilde{Y}_1^{n},M_4)$ is the decoded $(M_0,M_1,M_3)$ at receiver~1, and $(\check{M}_0,\check{M}_2,\check{M}_4)=g_2(\tilde{Y}_2^{n},M_3)$ is the decoded $(M_0,M_2,M_4)$ at receiver~2; $\tilde{Y}^n_{i}=(Y^n_{i},S^n),$ if the channel state is available at receiver~$i$, and $\tilde{Y}^n_{i}=Y^n_{i}$ otherwise.

A $\left(2^{nR_0},2^{nR_1},2^{nR_2},2^{nR_3},2^{nR_4},n\right)$ non-causal code for the channel consists of an encoding function,
\begin{align*}
f\hskip-1pt:\hskip-1pt\mathcal{M}_0\hskip-1pt\times\hskip-1pt \mathcal{M}_1\hskip-1pt \times\hskip-1pt \mathcal{M}_2\hskip-1pt \times\hskip-1pt \mathcal{M}_3\hskip-1pt \times\hskip-1pt \mathcal{M}_4\hskip-1pt \times\hskip-1pt {\mathcal{S}}^n\hskip-1pt\rightarrow\hskip-1pt \mathcal{X}^n,
\end{align*}
i.e., $X^n=f(M_0,M_1,M_2,M_3,M_4,S^n)$. This code also consists of two decoding functions which are defined the same as for the causal code. 

The average probability of error for a causal or a non-causal code is defined as
\vspace{-2pt}
\begin{multline*}
P_e^{(n)}=P((\hat{M}_0,\hat{M}_1,\hat{M}_3)\neq(M_0,M_1,M_3)\;\text{or}\; \\  (\check{M}_0,\check{M}_2,\check{M}_4)\neq(M_0,M_2,M_4)).
\end{multline*}

\begin{definition}
	For causal (non-causal) cases, a rate tuple $(R_0,R_1,R_2,R_3,R_4)$ is said to be \textit{achievable} for the channel if there exists a sequence of $\left(2^{nR_0},2^{nR_1},2^{nR_2},2^{nR_3},2^{nR_4},n\right)$ causal (non-causal) codes with $P_e^{(n)}\rightarrow 0$ as $n \rightarrow \infty$.
\end{definition}
\begin{definition}
	The \textit{capacity region} of the channel is the closure of the set of all achievable rate tuples $(R_0,R_1,R_2,R_3,R_4)$.
\end{definition}

\begin{definition}
	The two-receiver memoryless broadcast channel with states is said to be \textit{physically degraded} if $(X,S)\rightarrow Y_1\rightarrow Y_2$ form a Markov chain,
	and it is said to be \textit{stochastically degraded} or \textit{degraded} if there exists a $Y'_2$ such that $(X,S)\rightarrow Y_1\rightarrow Y'_2$ form a Markov chain, and $p_{_{Y_2\mid X,S}}\hspace{-2pt}\left(y_2\mid x,s\right)=p_{_{Y'_2\mid X,S}}\hspace{-2pt}\left(y_2\mid x,s\right)$
\end{definition}

\begin{figure*}[t]
	\setcounter{equation}{5}
	\begin{align}
		R_0+R_1&<I(U_0,U_1;\tilde{Y}_1)-I(U_0,U_1;S)\label{eq:marton1cssi},\\
		R_0+R_2&<I(U_0,U_2;\tilde{Y}_2)-I(U_0,U_2;S)\label{eq:marton2cssi},\\
		R_0+R_1+R_2&<I(U_0,U_1;\tilde{Y}_1)+I(U_2;\tilde{Y}_2\mid U_0)-I(U_1;U_2\mid U_0)-I(U_0,U_1,U_2;S)\label{eq:marton3cssi},\\
		R_0+R_1+R_2&<I(U_1;\tilde{Y}_1\mid U_0)+I(U_0,U_2;\tilde{Y}_2)-I(U_1;U_2\mid U_0)-I(U_0,U_1,U_2;S)\label{eq:marton4cssi},\\
		2R_0+R_1+R_2&<I(U_0,U_1;\tilde{Y}_1)+I(U_0,U_2;\tilde{Y}_2)-I(U_1;U_2\mid U_0)-I(U_0,U_1,U_2;S)-I(U_0;S)\label{eq:marton5cssi},
	\end{align}
	\hrulefill
	\vspace{-10pt}
\end{figure*}

\section{Broadcast Channel without RMSI}
In this section, we address the two-receiver memoryless broadcast channel with states, without RMSI, i.e., $(M_3,M_4)=(0,0)$. We first propose a transmission scheme and derive an inner bound for the causal category. We then present the best-known inner bound for the non-causal category~\cite[Theorem 2]{MUNonCausal}. We finally show that we can have a unified inner bound to cover both the causal and non-causal categories, despite the achievability schemes being different.

\subsection{With Causal CSIT}\label{Sec:SchemcausalwithoutRMSI}
We utilize Marton coding~\cite[p. 208]{NITBook}, superposition coding and Shannon strategy~\cite[p.\ 176]{NITBook} to construct a transmission scheme and derive an inner bound for the causal category, stated as Theorem~\ref{innerboundcausal}.

\begin{theorem}\label{innerboundcausal}
	A rate triple $(R_0,R_1,R_2)$ is achievable for the causal category if it satisfies
\vspace{-2pt}
\setcounter{equation}{0}
\begin{align}
 R_0+R_1&<I(U_0,U_1;\tilde{Y}_1)\label{eq:marton1causal},\\
 R_0+R_2&<I(U_0,U_2;\tilde{Y}_2)\label{eq:marton2causal},\\
 R_0+R_1+R_2&<I(U_0,U_1;\tilde{Y}_1)+I(U_2;\tilde{Y}_2\mid U_0)\nonumber\\
 &\hskip68pt-I(U_1;U_2\mid U_0),\label{eq:marton3causal}\\
 R_0+R_1+R_2&<I(U_1;\tilde{Y}_1\mid U_0)+I(U_0,U_2;\tilde{Y}_2)\nonumber\\
 &\hskip72pt-I(U_1;U_2\mid U_0),\label{eq:marton4causal}\\
 2R_0+R_1+R_2&<I(U_0,U_1;\tilde{Y}_1)+I(U_0,U_2;\tilde{Y}_2)\nonumber\\
 &\hskip70pt-I(U_1;U_2\mid U_0),\label{eq:marton5causal}
\end{align}
for some distribution $p(u_0,u_1,u_2)$ and some function $x=\gamma(u_0,u_1,u_2,\tilde{s})$. $\tilde{S}=S$ if the channel state is available causally at the transmitter, and $\tilde{S}=0$ if it is not available at the transmitter. $\tilde{Y}_i=(Y_i,S),\;i=1,2,$ if the channel state is available at receiver $i$, and $\tilde{Y}_i=Y_i$ otherwise.
\end{theorem}
\begin{remark}
	By using the random variable $\tilde{S}$, we address also the cases where the channel state (i) is not available at any node, or (ii) is not available at the transmitter and is available at either one receiver or both receivers.	
\end{remark}
\begin{remark}
	The inner bound in Theorem~\ref{innerboundcausal} is the best inner bound for the causal category without RMSI, and so it is tight for all the special cases whose capacity region is known~\cite{DegradedBCCausalNonCausal},~\cite[p. 184]{NITBook}.  	
\end{remark}
\begin{IEEEproof}
	[Proof of Theorem~\ref{innerboundcausal}](\textit{Codebook Construction}) The codebook of the transmission scheme is formed from three subcodebooks which are constructed according to the distribution $p(u_0,u_1,u_2)$. Before the subcodebook construction, using rate splitting, $M_i,\;\,i=1,2,$ is divided into the two independent messages $M_{i1}$ of rate $R_{i1}\geq0$, and $M_{i2}$ of rate $R_{i2}\geq0$ such that $R_i=R_{i1}+R_{i2}$. Subcodebook~0 consists of i.i.d. codewords
	\begin{align*}
		u_0^n(m_0,m_{11},m_{21}),
	\end{align*}
	generated according to $\prod_{j=1}^{n}p_{_{U_0}}(u_{0,j})$. Subcodebook~$i$, $i=1,2$, consists of codewords
	\begin{align*}
		u_i^n(m_0,m_{11},m_{21},m_{i2},l_i),
	\end{align*}
	generated according to 
	\begin{align*}
		\prod_{j=1}^{n}p_{_{U_i\hskip-1pt\mid\hskip-1pt U_0}}(u_{i,j}\hskip-2pt\mid\hskip-2pt u_{0,j}(m_0,m_{11},m_{21})),
	\end{align*} 
	where $l_i\hskip-2pt\in\hskip-2pt\{1,\ldots,2^{nR'_i}\}$, i.e., for each $u_0^n(m_0,m_{11},m_{21})$, $2^{n(R_{i2}+R'_i)}$ codewords are generated.
	
	(\textit{Encoding}) For the encoding, given $\{m_i\}_{i=0}^2$, we first find a pair $(l_1,l_2)$ such that 
	\begin{align*}
		\left(U_0^n\left(\cdot\right),U_1^n\left(\cdot,l_1\right),U_2^n\left(\cdot,l_2\right)\right)\in\mathcal{T}_{\epsilon'}^n,
	\end{align*}
	where $\mathcal{T}_{\epsilon'}^n$ is the set of jointly $\epsilon'$-typical $n$-sequences with respect to the considered distribution~\cite[p. 29]{NITBook}. If there is more than one pair, we arbitrary choose one of them, and if there does not exist one pair, we choose $(l_1,l_2)=(1,1)$. We then construct the transmitted codeword as $x_j=\gamma(u_{0,j}\left(\cdot\right),u_{1,j}\left(\cdot\right),u_{2,j}\left(\cdot\right),\tilde{s}_j),\;\,j=1,2,\ldots,n$; $\tilde{S}_j=S_j\;\;\forall j,$ if the channel state is available causally at the transmitter, and $\tilde{S}_j=0\;\;\forall j,$ if it is not available at the transmitter.
	
	(\textit{Decoding}) Receiver~1 decodes $\left(\hat{m}_0,\hat{m}_{11},\hat{m}_{12}\right)$, if it is the unique tuple that satisfies 
	\begin{align*}
		\left(U_0^n\left(\cdot\right),U_1^n\left(\cdot,l_1\right),\tilde{Y}_1^n \right)\in\mathcal{T}_{\epsilon_1}^n\;\,\text{ for some}\;\,m_{21}\,\text{and } l_1;
	\end{align*}
	otherwise an error is declared. Receiver~2 similarly decodes $\left(\check{m}_0,\check{m}_{21},\check{m}_{22}\right)$, if it is the unique tuple that satisfies
	\begin{align*}
		\left(U_0^n\left(\cdot\right),U_2^n\left(\cdot,l_2\right),\tilde{Y}_2^n\right)\in\mathcal{T}_{\epsilon_2}^n\;\,\text{ for some}\;\,m_{11}\,\text{and }l_2;
	\end{align*}
	otherwise an error is declared.
	
	To derive sufficient conditions for achievability, we assume without loss of generality the transmitted messages are each equal to one by the symmetry of code construction, and  $(l_1,l_2)=(l^*_1,l^*_2)$ where $1\leq l^*_i\leq 2^{nR'_i}$. Receiver~1 makes an error only if one or more of the following events happen.
	\begin{align*}
		\mathcal{E}_0&\hspace{-3pt}:\hspace{-3pt}\left(U_0^n\left(1,1,1\right),U_1^n\left(1,1,1,1,l_1\right),U_2^n\left(1,1,1,1,l_2\right)\right)\notin\mathcal{T}_{\epsilon'}^n\\
		&\hskip165pt\text{ for all }l_1\text{ and }l_2,\\
		\mathcal{E}_{11}&\hspace{-3pt}:\hspace{-3pt}\left(U_0^n\left(1,1,1\right),U_1^n(1,1,1,1,l^*_1),\tilde{Y}_1^n\right)\notin\mathcal{T}_{\epsilon_1}^n,\\
		\mathcal{E}_{12}&\hspace{-3pt}:\hspace{-3pt}\left(U_0^n\left(1,1,1\right),U_1^n\left(1,1,1,m_{12},l_1\right),\tilde{Y}_1^n\right)\in\mathcal{T}_{\epsilon_1}^n\\
		&\hskip125pt\text{ for some }m_{12}\neq1\text{ and }l_1,\\
		\mathcal{E}_{13}&\hspace{-3pt}:\hspace{-3pt}\left(U_0^n\left(1,1,m_{21}\right),U_1^n\left(1,1,m_{21},m_{12},l_1\right),\tilde{Y}_1^n\right)\hskip-3pt\in\hskip-2pt\mathcal{T}_{\epsilon_1}^n\\
		&\hskip85pt\text{ for some }m_{21}\neq1,m_{12}\neq1\text{ and }l_1,\\
		\mathcal{E}_{14}&\hspace{-3pt}:\hspace{-3pt}\left(\hspace{-3pt}U_0^n\left(m_0,m_{11},m_{21}\right)\hspace{-2pt},\hspace{-1pt}U_1^n\left(m_0,m_{11},m_{21},m_{12},l_1\right)\hspace{-2pt},\hspace{-2pt}\tilde{Y}_1^n\hspace{-3pt}\right)\hspace{-3pt}\in\hspace{-3pt}\mathcal{T}_{\epsilon_1}^n\nonumber\\
		&\hskip40pt\text{ for some }(m_{0},m_{11})\neq(1,1),m_{21},m_{12} \text{ and }l_1.
	\end{align*}
	Events leading to an error at receiver~2 are written similarly. Based on the error events, and using the packing lemma~\cite[p. 45]{NITBook} and the mutual covering lemma~\cite[p.~208]{NITBook}, sufficient conditions for achievability are
	\begin{align*}
		R'_1+R'_2&>I(U_1;U_2\mid U_0),\\
		R_{12}+R'_1&<I(U_1;\tilde{Y}_1\mid U_0),\\
		R_0+R_1+R_{21}+R'_1&<I(U_0,U_1;\tilde{Y}_1),\\
		R_{22}+R'_2&<I(U_2;\tilde{Y}_2\mid U_0),\\
		R_0+R_2+R_{11}+R'_2&<I(U_0,U_2;\tilde{Y}_2).
	\end{align*}
	We finally perform Fourier-Motzkin elimination to obtain the conditions in \eqref{eq:marton1causal}--\eqref{eq:marton5causal}.
\end{IEEEproof}

\begin{figure*}[t]
	\setcounter{equation}{10}
	\begin{align}
	R_0+R_1+R_3&<I(U_0,U_1;\tilde{Y}_1)-I(U_0,U_1;S)\label{eq:marton1cssirmsi},\\
	R_0+R_2+R_4&<I(U_0,U_2;\tilde{Y}_2)-I(U_0,U_2;S)\label{eq:marton2cssirmsi},\\
	R_0+R_1+R_2+R_3&<I(U_0,U_1;\tilde{Y}_1)+I(U_2;\tilde{Y}_2\mid U_0)-I(U_1;U_2\mid U_0)-I(U_0,U_1,U_2;S)\label{eq:marton3cssirmsi},\\
	R_0+R_1+R_2+R_4&<I(U_1;\tilde{Y}_1\mid U_0)+I(U_0,U_2;\tilde{Y}_2)-I(U_1;U_2\mid U_0)-I(U_0,U_1,U_2;S)\label{eq:marton4cssirmsi},\\
	2R_0+R_1+R_2+R_3+R_4&<I(U_0,U_1;\tilde{Y}_1)+I(U_0,U_2;\tilde{Y}_2)-I(U_1;U_2\mid U_0)-I(U_0,U_1,U_2;S)-I(U_0;S)\label{eq:marton5cssirmsi},
	\end{align}
	\vskip-0pt
	\hrulefill
	\vskip-0pt
\end{figure*}

\subsection{With Non-causal CSIT}\label{Sec:SchemnoncausalwithoutRMSI}
Marton coding, superposition coding, and Gelfand-Pinsker coding were used to derive an inner bound for the memoryless broadcast channel with non-causal CSIT~\cite[Theorem 2]{MUNonCausal}. This inner bound, stated as Proposition~\ref{innerboundnoncausal}, is the best-known inner bound for the non-causal category without RMSI, and so it is tight for all the special cases whose capacity region is known.

\begin{proposition}\label{innerboundnoncausal}
	 A rate triple $(R_0,R_1,R_2)$ is achievable for the non-causal category if it satisfies~\eqref{eq:marton1cssi}--\eqref{eq:marton5cssi} for some distribution $p(u_0,u_1,u_2\hspace{-2pt}\mid\hspace{-2pt}\tilde{s})$ and some function $x=\gamma(u_0,u_1,u_2,\tilde{s})$. $\tilde{S}=S$ if the channel state is available non-causally at the transmitter, and $\tilde{S}=0$ if it is not available at the transmitter.
\end{proposition}

Here, we review the scheme achieving this inner bound in order to highlight its differences with the scheme for the causal category. 

(\textit{Codebook Construction}) This inner bound is achieved using a transmission scheme formed from three subcodebooks. These subcodebooks are constructed according to the distribution $p(u_0,u_1,u_2\mid \tilde{s})$. Here $(U_0,U_1,U_2)$ is not independent of $S$ when the channel state is available at the transmitter as opposed to the scheme for the channel with causal CSIT. Subcodebook~0 consists of i.i.d. codewords
\begin{align*}
u_0^n(m_0,m_{11},m_{21},l_0)
\end{align*}
generated according to $\prod_{j=1}^{n}p_{_{U_0}}(u_{0,j})$ where $l_0\in\{1,\ldots,2^{nR'_0}\}$, i.e., for each $(m_0,m_{11},m_{21})$, $2^{nR'_0}$ codewords are generated. Subcodebook $i,\;\;i=1,2,$ consists of codewords
\begin{align*}
u_i^n(m_0,m_{11},m_{21},l_0,m_{i2},l_i)
\end{align*}
generated according to 
\begin{align*}
\prod_{j=1}^{n}p_{_{U_i\hskip-1pt\mid\hskip-1pt U_0}}(u_{i,j}\hskip-2pt\mid\hskip-2pt u_{0,j}(m_0,m_{11},m_{21},l_0)),
\end{align*}
where $l_i\hskip-2pt\in\hskip-2pt\{1,\ldots,2^{nR'_i}\}$.

(\textit{Encoding}) For the encoding, given $\{m_i\}_{i=0}^2$, we first find a triple $(l_0,l_1,l_2)$ such that 
\begin{align*}
\left(U_0^n\left(\cdot,l_0\right),U_1^n\left(\cdot,l_0,\cdot,l_1\right),U_2^n\left(\cdot,l_0,\cdot,l_2\right),\tilde{S}^n\right)\in\mathcal{T}_{\epsilon'}^n,
\end{align*}
where $\tilde{S}^n=(\tilde{S}_1,\tilde{S}_2,\ldots,\tilde{S}_n)$; $\tilde{S}_j=S_j\;\;\forall j,$ if the channel state is available non-causally at the transmitter, and $\tilde{S}_j=0\;\;\forall j,$ if it is not available at the transmitter. If there is more than one $(l_0,l_1,l_2)$, we arbitrary choose one of them, and if there does not exist one, we choose $(l_0,l_1,l_2)=(1,1,1)$. We then construct the transmitted codeword as $x_j=\gamma(u_{0,j}\left(\cdot\right),u_{1,j}\left(\cdot\right),u_{2,j}\left(\cdot\right),\tilde{s}_j),\;\,j=1,2,\ldots,n$.

(\textit{Decoding}) Receiver~1 decodes $\left(\hat{m}_0,\hat{m}_{11},\hat{m}_{12}\right)$, if it is the unique tuple that satisfies 
\begin{align*}
\left(U_0^n\left(\cdot,l_0\right),U_1^n\left(\cdot,l_0,\cdot,l_1\right),\tilde{Y}_1^n \right)\in\mathcal{T}_{\epsilon_1}^n\text{ for some }(m_{21},l_0,l_1);
\end{align*}
otherwise an error is declared. Receiver~2 decodes $\left(\check{m}_0,\check{m}_{21},\check{m}_{22}\right)$, if it is the unique tuple that satisfies 
\begin{align*}
\left(U_0^n\left(\cdot,l_0\right),U_2^n\left(\cdot,l_0,\cdot,l_2\right),\tilde{Y}_2^n\right)\in\mathcal{T}_{\epsilon_2}^n\text{ for some }(m_{11},l_0,l_2);
\end{align*}
otherwise an error is declared.

\subsection{A Unified Inner Bound}
Here we discuss that we can have a unified inner-bound expression for both the causal and non-causal categories. The inequalities in~\eqref{eq:marton1cssi}-\eqref{eq:marton5cssi} can be used for both the causal and non-causal categories. This is because, for the causal category, $(U_0,U_1,U_2)$ is independent of  $S$, and the terms $I(U_0;S)$, $I(U_0,U_1;S)$, $I(U_0,U_2;S)$ and $I(U_0,U_1,U_2;S)$ are zero. Then the inequalities in \eqref{eq:marton1cssi}-\eqref{eq:marton5cssi} reduce to the ones in \eqref{eq:marton1causal}-\eqref{eq:marton5causal}, and we can have a unified inner bound, stated as Corollary~\ref{unifiedwithoutRMSI}.
\begin{corollary}\label{unifiedwithoutRMSI}
A rate triple $(R_0,R_1,R_2)$ is achievable for the causal category if it satisfies \eqref{eq:marton1cssi}--\eqref{eq:marton5cssi} for some distribution $p(u_0,u_1,u_2)$ and some function $x=\gamma(u_0,u_1,u_2,\tilde{s})$; it is achievable for the non-causal category if it satisfies \eqref{eq:marton1cssi}--\eqref{eq:marton5cssi} for some distribution $p(u_0,u_1,u_2\mid \tilde{s})$ and some function $x=\gamma(u_0,u_1,u_2,\tilde{s})$.
\end{corollary}

The capacity region of a non-causal case is larger than or equal to the capacity region of the corresponding causal case. However, an inner bound for a non-causal case is not necessarily larger than an inner bound for the corresponding causal case when the scheme for the latter is not a special case of the scheme for the former. By having a unified inner bound, we can show that the inner bound for a non-causal case (Proposition~\ref{innerboundnoncausal}) is larger than or equal to the inner bound for the corresponding causal case (Theorem~\ref{innerboundcausal}). This is because the domain of the unified inner bound for a causal case (including all distributions $p(u_0,u_1,u_2)$ and all functions $x=\gamma(u_0,u_1,u_2,\tilde{s})$) is a subset of the domain of the unified inner bound for the corresponding non-causal case (including all distributions $p(u_0,u_1,u_2\mid \tilde{s})$ and all functions $x=\gamma(u_0,u_1,u_2,\tilde{s})$).

In Appendix A, we show that the scheme for the causal category (achieving the inner bound in Theorem~\ref{innerboundcausal}) is not a special case of the scheme for the non-causal category (achieving the inner bound in Proposition~\ref{innerboundnoncausal}). However, by considering only some special cases of its parameters, we show that the scheme for the non-causal category asymptomatically almost surely has the same codebook construction, encoding and decoding as the scheme for the causal category.

\section{Broadcast Channel with RMSI}\label{Section:CCSIatTransmitter}
In this section, we address the two-receiver memoryless broadcast channel with states and RMSI. We derive a unified inner bound that covers both categories with RMSI. This inner bound contains the unified inner bound in Corollary~\ref{unifiedwithoutRMSI} as a special case.

\subsection{Moving from Without RMSI to With RMSI}
To take the RMSI into account, we apply a pre-coding to the transmission schemes achieving the best inner bounds for the causal and non-causal categories without RMSI. This pre-coding was proposed in our previous work for the memoryless channel without state, with RMSI~\cite{TwoReceiverwithoutState}. In this pre-coding, $M_\text{m}=(M_0,M_3,M_4)$ is considered as a new common message, and only $M_1$ and $M_2$ are treated as the private messages. $M_\text{m}$, $M_1$ and $M_2$ are then fed to the transmission scheme of the channel without RMSI. Although receiver~1 need not decode $M_4$, having $M_4$ as a part of the common message does not impose any extra constraint; this is because receiver~1 knows $M_4$ a priori. The same argument applies to $M_3$ for receiver~2. Since receiver~1 knows $M_4$ a~priori, and receiver~2 knows $M_3$ a~priori, receiver~1 decodes $M_\text{m}$ over a set of $2^{n(R_0+R_3)}$ candidates, and receiver~2 decodes it over a set of $2^{n(R_0+R_4)}$ candidates.

\begin{conjecture}
	We conjecture that our pre-coding is an optimal pre-coding in the sense that if a scheme achieves the capacity region of a channel without RMSI, then the scheme, constructed by applying our pre-coding to that scheme, also achieves the capacity region of the same channel with RMSI.	
\end{conjecture}

\subsection{A Unified Inner Bound}
We here present our unified inner bound for the causal and non-causal categories with RMSI, stated as Theorem~\ref{unifiedwithRMSI}.
\begin{theorem}\label{unifiedwithRMSI}
	A rate tuple $(R_0,R_1,R_2,R_3,R_4)$ is achievable for the causal category with RMSI if it satisfies \eqref{eq:marton1cssirmsi}--\eqref{eq:marton5cssirmsi} for some distribution $p(u_0,u_1,u_2)$ and some function $x=\gamma(u_0,u_1,u_2,\tilde{s})$. It is achievable for the non-causal category with RMSI if it satisfies \eqref{eq:marton1cssirmsi}--\eqref{eq:marton5cssirmsi} for some distribution $p(u_0,u_1,u_2\mid \tilde{s})$ and some function $x=\gamma(u_0,u_1,u_2,\tilde{s})$.
\end{theorem}

\begin{remark}\label{remarkforwithRMSI}
	Having a unified inner bound for the causal and non-causal categories without RMSI, and applying the same pre-coding to both are the two basic reasons for having a unified inner bound for the causal and non-causal categories with RMSI.
\end{remark}

\begin{IEEEproof}[Proof of Theorem~\ref{unifiedwithRMSI}]We prove Theorem~\ref{unifiedwithRMSI} by applying our pre-coding to the causal and non-causal categories without RMSI.
	
	\subsubsection{With Causal CSIT}
	For this category, we apply our pre-coding to the transmission scheme for the causal category without RMSI, introduced in Section~\ref{Sec:SchemcausalwithoutRMSI}. Based on our method, Subcodebook~0 consists of i.i.d. codewords
	\begin{align*}
	u_0^n(m_0,m_3,m_4,m_{11},m_{21}),
	\end{align*}
	generated according to $\prod_{j=1}^{n}p_{_{U_0}}(u_{0,j})$. Subcodebook $i$, $i=1,2$, consists of codewords
	\begin{align*}
	u_i^n(m_0,m_3,m_4,m_{11},m_{21},m_{i2},l_i),
	\end{align*}
	generated according to 
	\begin{align*}
	\prod_{j=1}^{n}p_{_{U_i\hskip-1pt\mid\hskip-1pt U_0}}(u_{i,j}\hskip-2pt\mid\hskip-2pt u_{0,j}(m_0,m_3,m_4,m_{11},m_{21})),
	\end{align*} 
	where $l_i\hskip-2pt\in\hskip-2pt\{1,\ldots,2^{nR'_i}\}$.
	
	Encoding and decoding are performed similarly to the case without RMSI. For the encoding, given $\{m_i\}_{i=0}^4$, we first find a pair $(l_1,l_2)$ such that 
	\begin{align*}
	\left(U_0^n\left(\cdot\right),U_1^n\left(\cdot,l_1\right),U_2^n\left(\cdot,l_2\right)\right)\in\mathcal{T}_{\epsilon'}^n.
	\end{align*}
	If there does not exist one pair, we choose $(l_1,l_2)=(1,1)$. We then construct the transmitted codeword as $x_j=\gamma(u_{0,j}\left(\cdot\right),u_{1,j}\left(\cdot\right),u_{2,j}\left(\cdot\right),\tilde{s}_j),\;\,j=1,2,\ldots,n$.
	
	For the decoding, receiver~1 decodes $\left(\hat{m}_0,\hat{m}_{11},\hat{m}_{12},\hat{m}_3\right)$, if it is the unique tuple that satisfies 
	\begin{align*}
	\left(U_0^n\left(\cdot\right),U_1^n\left(\cdot,l_1\right),\tilde{Y}_1^n \right)\in\mathcal{T}_{\epsilon_1}^n\;\,\text{ for some}\;\,m_{21}\,\text{and } l_1;
	\end{align*}
	otherwise an error is declared. Since this receiver knows $M_4$ as side information, it decodes $(u_0^n,u_1^n)$ over a set of $2^{n(R_0+R_1+R_3+R_{21}+R'_1)}$ candidates. Receiver~2 decodes $\left(\check{m}_0,\check{m}_{21},\check{m}_{22},\check{m}_4\right)$, if it is the unique tuple that satisfies 
	\begin{align*}
	\left(U_0^n\left(\cdot\right),U_2^n\left(\cdot,l_2\right),\tilde{Y}_2^n\right)\in\mathcal{T}_{\epsilon_2}^n\;\,\text{ for some}\;\,m_{11}\,\text{and }l_2;
	\end{align*}
	otherwise an error is declared. Since this receiver knows $M_3$ as side information, it decodes $(u_0^n,u_2^n)$ over a set of $2^{n(R_0+R_2+R_4+R_{11}+R'_2)}$ candidates.
	
	Based on error events, written similarly to the case without RMSI, and using the packing lemma~\cite[p. 45]{NITBook} and the mutual covering lemma~\cite[p. 208]{NITBook}, sufficient conditions for achievability are
	\begin{align*}
	R'_1+R'_2&>I(U_1;U_2\mid U_0),\\
	R_{12}+R'_1&<I(U_1;\tilde{Y}_1\mid U_0),\\
	R_0+R_1+R_3+R_{21}+R'_1&<I(U_0,U_1;\tilde{Y}_1),\\
	R_{22}+R'_2&<I(U_2;\tilde{Y}_2\mid U_0),\\
	R_0+R_2+R_4+R_{11}+R'_2&<I(U_0,U_2;\tilde{Y}_2).
	\end{align*}
	After performing Fourier-Motzkin elimination, we obtain conditions \eqref{eq:marton1cssirmsi}--\eqref{eq:marton5cssirmsi}. Note that $(U_0,U_1,U_2)$ is independent of $S$ for this category. Then the terms $I(U_0;S)$, $I(U_0,U_1;S)$, $I(U_0,U_2;S)$, and $I(U_0,U_1,U_2;S)$ are zero. The inner bound is computed over all distributions $p(u_0,u_1,u_2)$ and all functions $x=\gamma(u_0,u_1,u_2,\tilde{s})$. 
	
	\subsubsection{With Non-causal CSIT}
	For this category, we apply our pre-coding to the transmission scheme for the non-causal category without RMSI, introduced in Section~\ref{Sec:SchemnoncausalwithoutRMSI}. The resulting changes to the codebook construction, encoding and decoding are similar to the ones for the causal category. Based on error events written similarly to the case without RMSI, and using the packing lemma and the multivariate covering lemma~\cite[p. 218]{NITBook}, sufficient conditions for achievability are
	\begin{align*}
	R'_0&>I(U_0;S),\\
	R'_0+R'_1&>I(U_0,U_1;S),\\
	R'_0+R'_2&>I(U_0,U_2;S),\\
	R'_0+R'_1+R'_2&>I(U_0,U_1,U_2;S)\\
	&\hskip30pt+I(U_1;U_2\mid U_0),\\
	R_{12}+R'_1&<I(U_1;\tilde{Y}_1\mid U_0),\\
	R_0+R_1+R_3+R_{21}+R'_0+R'_1&<I(U_0,U_1;\tilde{Y}_1),\\
	R_{22}+R'_2&<I(U_2;\tilde{Y}_2\mid U_0),\\
	R_0+R_2+R_4+R_{11}+R'_0+R'_2&<I(U_0,U_2;\tilde{Y}_2).
	\end{align*}
	After performing Fourier-Motzkin elimination, we obtain conditions \eqref{eq:marton1cssirmsi}--\eqref{eq:marton5cssirmsi}. Note that $(U_0,U_1,U_2)$ is not independent of $S$ for this category when the channel state is available at the transmitter. The inner bound is computed over all distributions $p(u_0,u_1,u_2\mid \tilde{s})$ and all functions $x=\gamma(u_0,u_1,u_2,\tilde{s})$.
\end{IEEEproof}

\section{New Capacity Results}\label{Section:DegradedChannel}
In this section, we present new capacity results for the two-receiver memoryless broadcast channel with RMSI. These results are established using our inner bound in Theorem~\ref{unifiedwithRMSI}. 

\subsection{With Causal CSIT and RMSI}
In this subsection, we present new capacity results for the causal category with RMSI, stated as Theorem~\ref{theorem:onlytransmitter}, and Theorem~\ref{theorem:transmitterandreceiver}.

\begin{theorem}\label{theorem:onlytransmitter}
The capacity region of the two-receiver degraded broadcast channel with RMSI where the channel state is available causally at only the transmitter, is the closure of the set of all rate tuples $(R_0,R_1,R_2,R_3,R_4)$, each satisfying
\begin{align*}
R_0+R_2+R_4&<I\left(U_0;Y_2\right),\\
R_0+R_1+R_2+R_3&<I\left(U_0,U_1;Y_1\right),\\
R_0+R_1+R_2+R_4&<I\left(U_0;Y_2\right)+I\left(U_1;Y_1\mid U_0\right),
\end{align*}
for some $p\left(u_0,u_1\right)$ and some function $x=\gamma\left(u_0,u_1,s\right)$.
\end{theorem}
We present the achievability proof in the following, and the converse proof in Appendix B.
\begin{IEEEproof}
(\textit{Achievability}) Achievability is proved by setting $U_2=0$ in \eqref{eq:marton1cssirmsi}--\eqref{eq:marton5cssirmsi}.	
\end{IEEEproof}
\begin{theorem}\label{theorem:transmitterandreceiver}
	The capacity region of the two-receiver degraded broadcast channel with RMSI where the channel state is available causally either at the transmitter and receiver~1 or at the transmitter and both receivers, is the closure of the set of all rate tuples $(R_0,R_1,R_2,R_3,R_4)$, each satisfying
	\begin{align*}
	R_0+R_2+R_4&<I(U_0;\tilde{Y}_2),\\
	R_0+R_1+R_2+R_3&<I(X;Y_1\mid S),\\
	R_0+R_1+R_2+R_4&<I(U_0;\tilde{Y}_2)+I\left(X;Y_1\mid U_0,S\right),
	\end{align*}
	for some $p\left(u_0,u_1\right)$ and some function $x=\gamma\left(u_0,u_1,s\right)$. $\tilde{Y}_2=(Y_2,S)$ if the channel state is available at receiver~2, and $\tilde{Y}_2=Y_2$  otherwise.
\end{theorem}
We present the achievability proof in the following, and the converse proof in Appendix C.
\begin{IEEEproof}
	(\textit{Achievability}) Achievability is proved by setting $U_2=0$ in \eqref{eq:marton1cssirmsi}--\eqref{eq:marton5cssirmsi}. Note that, for a causal case, $U_2=0$ implies that
	\begin{align*}
	I(U_0,U_1;Y_1,S)&=I(U_0,U_1;Y_1\mid S)=I(X;Y_1\mid S),\\
	I(U_1;Y_1,S\mid U_0)&\hspace{-0pt}=\hspace{-2pt}I(U_1;Y_1\mid U_0,S)\hspace{-2pt}=\hspace{-2pt}I(X;Y_1\mid U_0,S).
	\end{align*}	
\end{IEEEproof}

\subsection{With Non-causal CSIT and RMSI}
In this subsection, we first discuss that our inner bound in Theorem~\ref{unifiedwithRMSI} also achieves the capacity region of the case considered by Oechtering and Skoglund~\cite{BCNonCasualandSI0} (i.e., the memoryless broadcast channel when $(M_0,M_1,M_2)=(0,0,0)$ and the channel state is available non-causally at the transmitter and one of the receivers). This is because our transmission scheme for the non-causal category with RMSI reduces to the one used to establish the capacity region by setting $U_1=0$ and $U_2=0$. We also present new capacity results for the non-causal category with RMSI, stated as Theorem~\ref{theorem:noncausaldegraded}.
\begin{theorem}\label{theorem:noncausaldegraded}
	The capacity region of the two-receiver degraded broadcast channel with RMSI where the channel state is available non-causally either at the transmitter and receiver~1 or at the transmitter and both receivers, is the closure of the set of all rate tuples $(R_0,R_1,R_2,R_3,R_4)$, each satisfying
	\begin{align*}
		R_0+R_2+R_4&\hspace{-3pt}<\hspace{-3pt}I(U_0;\tilde{Y}_2)-I(U_0;S),\\
		R_0+R_1+R_2+R_3&\hspace{-3pt}<\hspace{-3pt}I(X;Y_1\mid S),\\
		R_0+R_1+R_2+R_4&\hspace{-3pt}<\hspace{-3pt}I(U_0;\tilde{Y}_2)\hspace{-2pt}-\hspace{-2pt}I(U_0;S)\hspace{-2pt}+\hspace{-2pt}I(X;Y_1\mid U_0,S),
	\end{align*}
	for some $p\left(u_0,u_1\mid s\right)$ and some function $x=\gamma\left(u_0,u_1,s\right)$. $\tilde{Y}_2=(Y_2,S)$ if the channel state is available at receiver~2, and $\tilde{Y}_2=Y_2$ otherwise.
\end{theorem} 
We present the achievability proof in the following, and the converse proof in Appendix C.
\begin{IEEEproof}
	(\textit{Achievability}) Achievability is proved by setting $U_2=0$ in \eqref{eq:marton1cssirmsi}--\eqref{eq:marton5cssirmsi}.	Note that, for a non-causal case, $U_2=0$ implies that $I(U_0,U_1;Y_1,S)\hspace{-2pt}-\hspace{-2pt}I(U_0,U_1;S)=I(X;Y_1\mid S)$.
\end{IEEEproof}

\section{Conclusion}
We first considered the two-receiver memoryless broadcast channel with states, without receiver message side information (RMSI). We addressed two categories of the channel: (i) channel with causal channel state side information at the transmitter (CSIT), and (ii) channel with non-causal CSIT. We proposed a transmission scheme and derived a general inner bound for the causal category. This inner bound is the best inner bound for this category. We also presented the best-known inner bound for the non-causal category. We showed that we can unify these two inner bounds to cover both the causal and non-causal categories. 

We then considered the two-receiver memoryless broadcast channel with states and RMSI where each receiver requests both common and private messages, and knows part of the private message requested by the other receiver as side information. We addressed the same two categories as for the channel without RMSI. We used a pre-coding to take the RMSI into account. We applied our pre-coding to the schemes achieving the best inner bounds for the two categories without RMSI. By this approach, we obtained a unified inner bound to cover both categories with RMSI. Using our inner bound, we also established new capacity results for a few special cases in both categories.

\section*{Appendix A}\label{Appendix:causalscheme-noncausalscheme}
Here we first show that the scheme for the causal category, described in Section~\ref{Sec:SchemcausalwithoutRMSI}, cannot be considered as a special case of the scheme for the non-causal category, described in Section~\ref{Sec:SchemnoncausalwithoutRMSI}. 
However, the rate regions achievable by both schemes have similar expression. This results in a unified inner bound for both causal and non-causal cases from which we can show that the inner bound for a non-causal case is at least as large as the the inner bound for  the corresponding causal case. We will explain the reason behind this observation despite them having different schemes.

We consider a scheme as a special case of another scheme when the latter reduces to the former by considering some special cases of its parameters, e.g., superposition coding is a special case of the scheme achieving Marton's inner bound with common message~\cite[p. 212]{NITBook}.

Consider the encoding rule of the scheme for the non-causal category where the encoder finds a $(U_0^n,U_1^n,U_2^n)$ such that
\begin{align*}
	\left(U_0^n,U_1^n,U_2^n,S^n\right)\in\mathcal{T}_{\epsilon'}^n(U_0,U_1,U_2,S).
\end{align*}
The encoder for the causal category cannot check this rule since this encoder only knows $S^n$ at the end of the transmission. Hence, the scheme for the causal category is not a special case of the scheme for the non-causal category.

By choosing $p(u_0,u_1,u_2\mid s)=p(u_0,u_1,u_2)$ for all $s$ and setting $R'_0=0$, the scheme for the non-causal category has the same codebook construction and decoding approach as the scheme for the causal category. The only difference is that, for the channel state realization $s^n$, the encoder for the non-causal category finds a $(u_0^n,u_1^n,u_2^n)$ such that
\begin{align*}
	(u_0^n,u_1^n,u_2^n)\in\mathcal{T}_{\epsilon'}^n(U_0,U_1,U_2\mid s^n),
\end{align*}
where 
\begin{multline*}
	\mathcal{T}_{\epsilon'}^n(U_0,U_1,U_2\mid s^n)=\\
	\{(u_0^n,u_1^n,u_2^n)\mid(u_0^n,u_1^n,u_2^n,s^n)\in\mathcal{T}_{\epsilon'}^n(U_0,U_1,U_2,S)\}.
\end{multline*}
and the encoder for the causal category finds a $(u_0^n,u_1^n,u_2^n)$ such that
\begin{align*}
	(u_0^n,u_1^n,u_2^n)\in \mathcal{T}_{\epsilon'}^n(U_0,U_1,U_2).
\end{align*}
So the transmitted codewords may be different. However, according to the properties of joint typicality \cite[p. 27]{NITBook}, we have
\begin{align*}
	\mathcal{T}_{\epsilon'}^n(U_0,U_1,U_2\mid s^n)\subseteq \mathcal{T}_{\epsilon'}^n(U_0,U_1,U_2),
\end{align*}
where the equality holds if $s^n\in\mathcal{T}_{\epsilon'}^n(S)$. Since 
\begin{align*}
	P\left(S^n\in\mathcal{T}_{\epsilon'}^n(S)\right)\rightarrow 1,
\end{align*}
as $n$ tends to infinity then
\begin{align*}
	P\left(\mathcal{T}_{\epsilon'}^n(U_0,U_1,U_2\mid S^n)= \mathcal{T}_{\epsilon'}^n(U_0,U_1,U_2)\right)\rightarrow 1,
\end{align*}
as $n$ tends to infinity. Hence, by choosing $p(u_0,u_1,u_2\mid s)=p(u_0,u_1,u_2)$, the scheme for the non-causal category asymptomatically almost surely has the same encoding as the scheme for the causal category. This leads the scheme for the non-causal category to achieve the same rate region as the scheme for the causal category.

\section*{Appendix B}\label{Appendix:causalonlytransmitter}
In this section, we present the converse proof of Theorem~\ref{theorem:onlytransmitter}. In the converse, we assume that the broadcast channel is physically degraded as the capacity region of the stochastically degraded broadcast channel is equal to its equivalent physically degraded broadcast channel.

\begin{IEEEproof}
	(\textit{Converse}) By Fano's inequality \cite[p. 19]{NITBook}, we have
	\begin{align}
	H\left(M_0,M_1,M_3\mid Y_1^n,M_4\right)\leq n\epsilon_{1,n}\label{eq:fano1},\\
	H\left(M_0,M_2,M_4\mid Y_2^n,M_3\right)\leq n\epsilon_{2,n}, \label{eq:fano2}
	\end{align}
	where $\epsilon_{i,n}\rightarrow 0$ as $n\rightarrow \infty$ for $i=1,2$. For the sake of simplicity, we use $\epsilon_{n}$ instead of $\epsilon_{i,n}$ for the remainder.
	From \eqref{eq:fano2} and the physically degradedness of the channel, we have
	\begin{align}
		H\left(M_0,M_2,M_4\mid Y_1^n,M_3\right)&\leq
		H\left(M_0,M_2,M_4\mid Y_2^n,M_3\right)\nonumber\\
		&\leq n{\epsilon_{n}},\label{fano23}
	\end{align}
	and from \eqref{eq:fano1} and \eqref{fano23}, we have
	\begin{align}
		H\left(M_0,M_2,M_3\mid Y_1^n,M_4\right)\leq 2n\epsilon_n.\label{fano24}
	\end{align}
	Using \eqref{eq:fano1}, \eqref{eq:fano2}, and \eqref{fano24}, we obtain the following necessary conditions for achievability
	\begin{align}
		nR_1 &\leq I(M_1;Y_1^n\mid M_0,M_2,M_3,M_4)+n\epsilon_n,\label{converse21}\\
		n(R_0+R_2+R_3) & \leq I(M_0,M_2,M_3;Y_1^n\mid M_4)+2n\epsilon_n,\label{converse22}\\
		n(R_0+R_2+R_4) & \leq I(M_0,M_2,M_4;Y_2^n\mid M_3)+n\epsilon_n.\label{converse23}
	\end{align}
	
	We now define the auxiliary random variables $U_{0,i}$ and $U_{1,i}$~as 
	\begin{align*}
		U_{0,i}&=(M_0,M_2,M_3,M_4,Y_1^{i-1}),\\
		U_{1,i}&=(M_1,S^{i-1}),
	\end{align*}
	and expand the mutual information terms in \eqref{converse21}--\eqref{converse23} respectively as follows.
	
	\begin{align}
		&\hskip-15pt I\left(M_1;Y_1^n\mid M_0,M_2,M_3,M_4\right)\nonumber\\
		&=\sum_{i=1}^{n}I\left(M_1;Y_{1,i}\mid M_0,M_2,M_3,M_4,Y_1^{i-1}\right)\nonumber\\
		&\leq\sum_{i=1}^{n}I\left(M_1,S^{i-1};Y_{1,i}\mid M_0,M_2,M_3,M_4,Y_1^{i-1}\right)\nonumber\\
		&=\sum_{i=1}^{n}I\left(U_{1,i};Y_{1,i}\mid U_{0,i}\right),\label{converse211}
	\end{align}
	
	\begin{align}
		&\hskip-35ptI\left(M_0,M_2,M_3;Y_1^n\mid M_4\right)\nonumber\\
		&=\sum_{i=1}^{n}I\left(M_0,M_2,M_3;Y_{1,i}\mid M_4,Y_1^{i-1}\right)\nonumber\\
		&\leq\sum_{i=1}^{n}I\left(M_0,M_2,M_3,M_4,Y_1^{i-1};Y_{1,i}\right)\nonumber\\
		&=\sum_{i=1}^{n}I\left(U_{0,i};Y_{1,i}\right),\label{converse221}
	\end{align}
	and
	\begin{align}
		&\hskip-20ptI\left(M_0,M_2,M_4;Y_2^n\mid M_3\right)\nonumber\\
		&=\sum_{i=1}^{n}I\left(M_0,M_2,M_4;Y_{2,i}\mid M_3,Y_{2}^{i-1}\right)\nonumber\\
		&\leq\sum_{i=1}^{n}I\left(M_0,M_2,M_3,M_4,Y_2^{i-1};Y_{2,i}\right)\nonumber\\
		&\leq\sum_{i=1}^{n}I\left(M_0,M_2,M_3,M_4,Y_2^{i-1},Y_1^{i-1};Y_{2,i}\right)\nonumber\\
		&\overset{(a)}{=}\sum_{i=1}^{n}I\left(M_0,M_2,M_3,M_4,Y_1^{i-1};Y_{2,i}\right)\nonumber\\
		&=\sum_{i=1}^{n}I\left(U_{0,i};Y_{2,i}\right),\label{converse231}
	\end{align}
	\\
	where $(a)$ follows from the physically degradedness of the channel. 
	
	Finally, since $\epsilon_n\rightarrow 0$ as $n\rightarrow \infty$, substituting \eqref{converse211}--\eqref{converse231} into \eqref{converse21}--\eqref{converse23}, and using the standard time-sharing argument~\cite[p. 114]{NITBook} complete the converse proof. Note that $(U_{0,i},U_{1,i})$ is independent of $S_i$, and $X_i$ is a function of $(U_{0,i},U_{1,i},S_i)$. 
\end{IEEEproof}

\section*{Appendix C}\label{Appendix:causaltransmitterandreceiver1}
In this section, we present the converse proof of Theorem~\ref{theorem:transmitterandreceiver} and Theorem~\ref{theorem:noncausaldegraded}. We here also assume that the broadcast channel is physically degraded.  
\begin{IEEEproof}
	(\textit{Converse})	By Fano's inequality, we have
	\begin{align}
		H(M_0,M_1,M_3\mid Y_1^n,S^n,M_4)\leq n{\epsilon_{n}},\label{fano31}\\
		H(M_0,M_2,M_4\mid \tilde{Y}_2^n,M_3)\leq n{\epsilon_{n}}\label{fano32}.
	\end{align}
	From \eqref{fano32} and the physically degradedness of the channel, we have
	\begin{multline}
		H\left(M_0,M_2,M_4\mid Y_1^n,S^n,M_3\right)\leq\\
		H\left(M_0,M_2,M_4\mid Y_2^n,S^n,M_3\right)\leq n{\epsilon_{n}}\label{fano33},
	\end{multline}
	and from \eqref{fano31} and \eqref{fano33}, we have
	\begin{align}
		H\left(M_0,M_1,M_2,M_3\mid Y_1^n,S^n,M_4\right)&\leq 2n\epsilon_n.\label{fano34}
	\end{align}
	Using \eqref{fano31}, \eqref{fano32}, and \eqref{fano34}, if a rate tuple $(R_0,R_1,R_2,R_3,R_4)$ is achievable, then it must satisfy
	\begin{align}
		nR_1  &\leq I(M_1;Y_1^n\mid M_0,M_2,M_3,M_4,S^n)\nonumber\\
		&\hskip40pt+n\epsilon_n,\label{converse31}\\
		n(R_0+R_1+R_2+R_3)&\leq I(M_0,M_1,M_2,M_3;Y_1^n\mid M_4,S^n)\nonumber\\
		&\hskip40pt+2n\epsilon_n,\label{converse32}\\
		n(R_0+R_2+R_4)&\leq I(M_0,M_2,M_4;\tilde{Y}_2^n\mid M_3)+n\epsilon_n.\label{converse33}
	\end{align}
	
	We now define the auxiliary random variables $U_{0,i}$ and $U_{1,i}$~as	
	\begin{align*}
		U_{0,i}&=(M_0,M_2,M_3,M_4,S^{i-1},S_{i+1}^n,Y_2^{i-1}),\\
		U_{1,i}&=(M_1,Y_1^{i-1}),
	\end{align*}
	and expand the mutual information terms in \eqref{converse31}--\eqref{converse33} respectively as follows.
	\begin{align}
		&\hskip-10pt I\left(M_1;Y_1^n\mid M_0,M_2,M_3,M_4,S^n\right)\nonumber\\
		&=\sum_{i=1}^{n}I\left(M_1;Y_{1,i}\mid M_0,M_2,M_3,M_4,S^n,Y_1^{i-1}\right)\nonumber\\
		&\overset{(a)}{=}\sum_{i=1}^{n}I\left(M_1;Y_{1,i}\mid M_0,M_2,M_3,M_4,S^n,Y_1^{i-1},Y_2^{i-1}\right)\nonumber\\
		&\leq\sum_{i=1}^{n}I\left(M_1,Y_1^{i-1};Y_{1,i}\mid M_0,M_2,M_3,M_4,S^n,Y_2^{i-1}\right)\nonumber\\
		&=\sum_{i=1}^{n}I\left(U_{1,i};Y_{1,i}\mid U_{0,i},S_i\right)\nonumber\\
		&=\sum_{i=1}^{n}I\left(X_i;Y_{1,i}\mid U_{0,i},S_i\right),\label{converse311}
	\end{align}
	
	\begin{align}
		&\hskip-10pt I\left(M_0,M_1,M_2,M_3;Y_1^n\mid M_4,S^n\right)\nonumber\\
		&=\sum_{i=1}^{n}I\left(M_0,M_1,M_2,M_3;Y_{1,i}\mid M_4,S^n,Y_1^{i-1}\right)\nonumber\\
		&\overset{(b)}{=}\sum_{i=1}^{n}I\left(M_0,M_1,M_2,M_3;Y_{1,i}\mid M_4,S^n,Y_1^{i-1},Y_2^{i-1}\right)\nonumber\\
		&\leq\sum_{i=1}^{n}I\left(U_{0,i},U_{1,i};Y_{1,i}\mid S_i\right)\nonumber\\
		&=\sum_{i=1}^{n}I\left(X_i;Y_{1,i}\mid S_i\right),\label{converse321}
	\end{align}
	\\
	and
	\\
	\begin{align}
		&\hskip-10ptI(M_0,M_2,M_4;\tilde{Y}_2^n\mid M_3)\nonumber\\
		&=\sum_{i=1}^{n}I(M_0,M_2,M_4;\tilde{Y}_{2,i}\mid M_3,\tilde{Y}_{2}^{i-1})\nonumber\\
		&\leq\sum_{i=1}^{n}I(M_0,M_2,M_3,M_4,S^{i-1},\tilde{Y}_2^{i-1};\tilde{Y}_{2,i})\nonumber\\
		&=\sum_{i=1}^{n}I(M_0,M_2,M_3,M_4,S^{i-1},S_{i+1}^n,\tilde{Y}_2^{i-1};\tilde{Y}_{2,i})\nonumber\\
		&\hskip40pt-I(S_{i+1}^n;\tilde{Y}_{2,i}\mid M_0,M_2,M_3,M_4,S^{i-1},\tilde{Y}_2^{i-1})\nonumber\\
		&\overset{(c)}{=}\sum_{i=1}^{n}I(M_0,M_2,M_3,M_4,S^{i-1},S_{i+1}^n,\tilde{Y}_2^{i-1};\tilde{Y}_{2,i})\nonumber\\
		&\hskip40pt-I(\tilde{Y}_2^{i-1};S_i\mid M_0,M_2,M_3,M_4,S^{i-1},S_{i+1}^n)\nonumber\\
		&\overset{(d)}{=}\sum_{i=1}^{n}I(M_0,M_2,M_3,M_4,S^{i-1},S_{i+1}^n,\tilde{Y}_2^{i-1};\tilde{Y}_{2,i})\nonumber\\
		&\hskip40pt-I(M_0,M_2,M_3,M_4,S^{i-1},S_{i+1}^n,\tilde{Y}_2^{i-1};S_i)\nonumber\\
		&=\sum_{i=1}^{n}I(U_{0,i};\tilde{Y}_{2,i})-I(U_{0,i};S_i),\label{converse331}
	\end{align}
	where $(a)$ and $(b)$ follow from the physically degradedness of the channel, $(c)$ from the Csisz\'{a}r sum identity~\cite[p. 25]{NITBook}, and $(d)$ from the independence of $(M_0,M_2,M_3,M_4,S^{i-1},S_{i+1}^n)$ and $S_i$ . Note that, for causal cases, $(U_{0,i},U_{1,i})$ is independent of $S_i$, but for non-causal cases  $(U_{0,i},U_{1,i})$ and $S_i$ are dependent. For both causal and non-causal cases, $X_i$ is a function of $(U_{0,i},U_{1,i},S_i)$.
	
	Finally, since $\epsilon_n\rightarrow 0$ as $n\rightarrow \infty$, substituting \eqref{converse311}--\eqref{converse331} into \eqref{converse31}--\eqref{converse33}, and using the standard time sharing argument complete the converse proof. 
\end{IEEEproof}

\bibliographystyle{IEEEtran}

\end{document}